\documentclass[12pt,a4paper]{article}
\pdfoutput=1
\usepackage[utf8]{inputenc}
\usepackage[english]{babel}
\usepackage{amsmath}
\usepackage{amsfonts}
\usepackage{amssymb}
\usepackage{graphicx}
\usepackage{color}
\usepackage{transparent}
\usepackage{fullpage}

\date{}
\author{ }
\title{}

\begin{document}

\def\papertitlepage{\baselineskip 3.5ex \thispagestyle{empty}}
\def\preprinumber#1#2{\hfill
\begin{minipage}{1.0in}
#1 \par\noindent #2
\end{minipage}}

\papertitlepage
\preprinumber{YITP-12-82}{} 
\vskip 2ex

\begin{center}

{\LARGE\bf\mathversion{bold}
On recursion relations in topological string theory}

\vspace{+30pt}

\end{center}

\baselineskip=3.5ex

\begin{center}
  Andrea Prudenziati\\

{\small
\vskip 3ex
{\it Yukawa Institute for Theoretical Physics, Kyoto University}\\[1ex]
{\it Kyoto 606-8502, Japan}\\
\vskip 1ex
{\tt prude@yukawa.kyoto-u.ac.jp }

}
\end{center}

\baselineskip=3.5ex

\vspace{+10pt}

\begin{abstract}

We discuss a link between the topological recursion relations derived algebraically by Witten and the holomorphic anomaly equation of Bershadsky, Cecotti, Ooguri and Vafa. This is obtained through the definition of an operator ${\cal{W}}_s$ that reproduces the recursion relations for topological string theory coupled to worldsheet gravity a la BCOV. This operator is contained inside an algebra that generalizes the tt* equations and whose direct consequence is the holomorphic anomaly equation itself.

\end{abstract}



\vspace{+20pt}

\section*{Introduction}

In \cite{Witten:1989ig}, \cite{Witten:1990hr}  Witten defined a class of two dimensional topological theories of gravity coupled with matter, whose distinctive feature is that the correlation functions are equal to an integral over the moduli space with the volume form expressed as the wedge product of certain differential forms associated to the each of the operators. In particular the explicit physical realization of this is that every operator contains an integer power of a scalar field $\phi$, the last component of the gravitational multiplet, whose value defines the so called gravitational degree;  in the moduli space integral this corresponds to the power of the two dimensional differential form $\alpha$ associated to the scalar $\phi$. Carefully evaluating the zeros of a section representing one of the $\alpha$'s, and if these have support only on the boundary of the moduli space of the Riemann surface, allows us to express a correlation function in terms of correlation functions containing a lower total gravitational degree and with genus lower or equal to the original one. The result are called topological recursion relations.

Indeed it is known that the holomorphic limit of the A model correlation functions basically obeys the Witten's abstract definition of topological gravity coupled to a two dimensional sigma model, \cite{Bershadsky:1993ta}. Moreover it computes Gromov-Witten invariants,  and these satisfy topological recursion relations. However A and B  topological string models are physically a different story from the theories considered by Witten: the target space is fixed to be a Calabi Yau threefold, in which case the gravitational multiplet scalar $\phi$ simply decouples ( with the only exception of the dilaton ). The coupling to two dimensional gravity comes instead from the analogy with the bosonic string theory construction of the gauge fixed path integral, and correlation functions are defined to contain a copy of the left and right moving twisted spin two supercurrents $G^-, \bar{G}^-$ for each of the moduli of the Riemann surface $\Sigma_{g,n}$ \footnote{genus $g$ and $n$ operator insertions}.

This way of defining topological string amplitudes led the authors of \cite{Bershadsky:1993cx} to the derivation of a different set of recursion relations expressing the antiholomorphic target space moduli derivative of correlation functions in terms of covariant holomorphic derivatives of amplitudes with either a lower genus or a lower number of operator insertions, or both. 
These relations are known as the holomorphic anomaly equation ( H.A.E. for short ). 

It is thus a natural question if it exists or not a formalism able to provide a direct connection between the topological recursion relations and the H.A.E. To my knowledge a similar issue was investigated so far only in \cite{Gomez:1994qk}, where the tt* equations  were derived from consistency conditions over an enlarged set of operator contact term algebra containing also the antiholomorphic sector, and the H.A.E. was obtained from the recursion relations written by Verlinde and Verlinde in \cite{Verlinde:1990ku} . Their assumptions, development and results are however quite different from the present discussion.

The strategy of the work will be simple. We want to formally apply the Witten's recursion relations machinery to  topological string theory correlation functions by roughly looking at $G^-, \bar{G}^-$ as a substitute of the field $\phi$. In fact we will see that these objects are, for certain properties, the analogue of what $\phi$ is for the gravitational descendants. This will lead to the definition of an operator ${\cal{W}}_s$ acting on correlation functions and producing what would have been the right hand side of the topological recursion relations if $\phi$ had been replaced by $G^-, \bar{G}^-$. The result will be that, imposing a certain commutator between ${\cal{W}}_s$ and the flat antiholomorphic derivative $\bar{\nabla}_{\bar{i}}$ ( from the tt* equations of \cite{Cecotti:1991me} ), is equivalent to the  H.A.E. In fact the full algebra we will consider is a generalization of the tt* equations and reads:

\[
[{\cal{W}}_s,\bar{\nabla}_{\bar{i}}] = \bar{\nabla}_{\bar{i}} 
\]
\[
[{\cal{W}}_s,\nabla_{i}] = 0
\]
\[
[\nabla_{i},\bar{\nabla}_{\bar{j}}] = [\nabla_{i},\nabla_j] = [\bar{\nabla}_{\bar{i}},\bar{\nabla}_{\bar{j}}]=0
\]

The first section reviews some background and establishes the notation we will use. The second and third derive the result for genus zero and one. The fourth further analyses the algebra between ${\cal{W}}_s$ and the flat derivatives and finally the last section considers the generalization at higher genus. Then we present some conclusions.

\section{General background}

In \cite{Witten:1989ig}, \cite{Witten:1990hr} two powerful relations were derived involving genus zero and one correlation functions for a wide class of 2 dimensional topological gravitational theories coupled to matter. These theories are defined by the requirement that the correlation functions in some point in the moduli space are expressed as an integral  of a volume form given by the wedge product of certain two dimensional cohomological forms associated to the operators. These operators are generically represented as ${\cal{O}}_{d,\alpha}$ and called the degree $d$ gravitational descendants of the matter operators ${\cal{O}}_{\alpha} = {\cal{O}}_{0,\alpha}$ belonging to the sigma model topological field theories described in \cite{Witten:1991zz}. 

\begin{equation}\label{c}
\langle {\cal{O}}_{d_{1},\alpha_{1}}\dots{\cal{O}}_{d_{n},\alpha_{n}}\rangle_{g} = \int_{{\cal{M}}_{g,n}}\hspace{-0,3cm} \alpha_{(1)}^{d_{1}}\wedge\dots\wedge \alpha_{(n)}^{d_{n}}\int_{{\cal{L}}_{g}}\hspace{-0,1cm}\beta_{(1)}\wedge \dots\wedge\beta_{(n)}
\end{equation}

Here ${\cal{M}}_{g,n}$ refers to the moduli space of the Riemann surface $\Sigma_{g,n}$, and ${\cal{L}}_{g}$ to the instanton moduli space of maps $X: \Sigma\rightarrow M$, $M$ being the target space.  Moreover $\alpha_{(i)}^{d_i} = \alpha_{(i)} \wedge \dots \wedge \alpha_{(i)}$ \footnote{we here slightly change the definition from the Witten's one discarding the $d_i!$ factor appearing in \cite{Witten:1989ig}, \cite{Witten:1990hr}. } with $ \alpha_{(i)} $ the first Chern class of $T^{*}\Sigma_{g,n}|_{\sigma_i}$, the cotangent space to $\Sigma_{g,n}$ at the position $\sigma_i$ of  ${\cal{O}}_{d_{i},\alpha_{i}}$; finally being $M_i$ the submanifold of $M$  Poincare dual to ( the differential form associated to ) the operator ${\cal{O}}_{\alpha_{i}}$, then $\beta_{(i)}$ is defined to be the Poincare dual to the submanifold of ${\cal{L}}_{g}$ determined by the condition $X(\sigma_i)\in M_i$. 

From the above definition and the complex dimension of ${\cal{M}}_{g,n}$ the selection rule   

\begin{equation}\label{c1}
\sum_{i=1}^n (d_i -1) = 3g-3
\end{equation}

is derived.

In this setup Witten obtained the following expressions for genus zero and one amplitudes, in the form of recursion relations:

\begin{equation}\label{a}
\langle {\cal{O}}_{d_{1},\alpha_{1}}\dots{\cal{O}}_{d_{n},\alpha_{n}}\rangle_{0} = 
\end{equation}
\[
= \hspace{-0,3cm}\sum_{X\cup Y=\{2,\dots,n-2\}}\langle {\cal{O}}_{d_{1}-1,\alpha_{1}}\prod_{r\in X}{\cal{O}}_{d_{r},\alpha_{r}}{\cal{O}}_{\alpha}\rangle_{0}\eta^{\alpha\beta}\langle {\cal{O}}_{\beta}\prod_{t\in Y}{\cal{O}}_{d_{t},\alpha_{t}}{\cal{O}}_{d_{n-1},\alpha_{n-1}}{\cal{O}}_{d_{n},\alpha_{n}}\rangle_{0}
\]

and

\begin{equation}\label{b}
\langle {\cal{O}}_{d_{1},\alpha_{1}}\dots{\cal{O}}_{d_{n},\alpha_{n}}\rangle_{1} = 
\end{equation}
\[
\frac{1}{12}\langle {\cal{O}}_{d_{1}-1,\alpha_{1}}\dots{\cal{O}}_{d_{n},\alpha_{n}}{\cal{O}}_{\alpha}{\cal{O}}_{\beta}\rangle_{0}\eta^{\alpha\beta}\hspace{+0,1cm} + \hspace{-0,2cm}\sum_{X\cup Y=\{2,\dots,n\}}\hspace{-0,2cm}\langle {\cal{O}}_{d_{1}-1,\alpha_{1}}\prod_{r\in X}{\cal{O}}_{d_{r},\alpha_{r}}{\cal{O}}_{\alpha}\rangle_{0}\eta^{\alpha\beta}\langle {\cal{O}}_{\beta}\prod_{t\in Y}{\cal{O}}_{d_{t},\alpha_{t}}\rangle_{1}
\]

Here $\eta_{\alpha\beta}$ is a metric on the space of operators whose definition given in \cite{Witten:1989ig} and \cite{Witten:1990hr} coincides with the one of \cite{Cecotti:1991me}, that we will soon review.

If we write the gravitational multiplet of the two dimensional theory as $(w_{\mu},\psi_{\mu},\phi)$, the  associated ghost number is respectively $(0,1,2)$ and the transformation rules under the BRST-like topological charge $Q$ are:

\[
\delta w_{\sigma} = i\epsilon\psi_{\sigma} 
\]
\begin{equation}\label{b11}
\delta \psi_{\sigma} = -\epsilon\partial_{\sigma}\phi 
\end{equation}
\[
\delta \phi = 0
\]

 Then the explicit field realization used by Witten for ${\cal{O}}_{d,\alpha}$, given ${\cal{O}}_{\alpha}$, is:  

\begin{equation}\label{b1}
{\cal{O}}_{d,\alpha} = {\cal{O}}_{\alpha}\phi^d
\end{equation}

From the above equations the topological charge has positive ghost number $+1$ and ${\cal{O}}_{d,\alpha}$ has ghost number $2d$; the solutions ${\cal{O}}_{d,\alpha}^{(1)}$ and $ {\cal{O}}_{d,\alpha}^{(2)}$ to the descent equation \footnote{the $d$ appearing in $d{\cal{O}}_{d,\alpha}$ below and in $dw$ in the next equation is the de Rham two dimensional differential, not the integer}

\[
0 = [Q,{\cal{O}}_{d,\alpha}]
\]
\begin{equation}\label{b1a}
d{\cal{O}}_{d,\alpha} = [Q,{\cal{O}}^{(1)}_{d,\alpha}]
\end{equation}
\[
d{\cal{O}}_{d,\alpha}^{(1)} = [Q,{\cal{O}}_{d,\alpha}^{(2)}]
\]
\[
d{\cal{O}}_{d,\alpha}^{(2)} = 0
\]

instead have decreasing ghost number $2d-1$ and $2d-2$ and explicit form ( for ${\cal{O}}_{\alpha}= {\cal{O}}_{0} = 1$ )

\[
{\cal{O}}_{d,0} = \phi^d \;\;\; {\cal{O}}_{d,0}^{(1)} = d \;\psi \;\phi^{d-1} \;\;\;{\cal{O}}_{d,0}^{(2)} = d \;dw\; \phi^{d-1} + 1/2\; d(d-1)\;\psi\wedge\psi \;\phi^{d-2}
\]

The starting point of the present paper is to formally translate the recursion relations derived in the above formalism to amplitudes in topological string theory of "BCOV" type defined as in \cite{Bershadsky:1993cx} . The naive justification for this is that, even if basically different in construction, BCOV type amplitudes substantially satisfy, in the holomorphic limit, all the important requirement of coupling of two dimensional gravity to topological theories of matter that were used by Witten in the derivation of the recursion relations ( look for example at the discussion in \cite{Bershadsky:1993ta} ). This is explicit for the A model computed Gromov-Witten invariants for which recursion relations are well known.

We begin with genus zero and $n$ marginal operator insertions. The Riemann surface moduli space integral is saturated by $\oint_{C_{\sigma}}G^{-}\oint_{C'_{\sigma}} \bar{G}^{-}$ around the position of certain operators, with  $G^{-},\bar{G}^{-}$ conventionally chosen to be the left and right moving spin two supercurrents associated to the spin one antitopological charge \footnote{the discussion is general and does not depend on the specific choice for the topological twist.}. Using a notation analogous to the case with gravitational descendants we define

\begin{equation}\label{c11}
\langle {\cal{O}}_{1,\alpha_{1}}\dots {\cal{O}}_{1,\alpha_{n-3}}{\cal{O}}_{\alpha_{n-2}}{\cal{O}}_{\alpha_{n-1}}{\cal{O}}_{\alpha_{n}}\rangle_0 \equiv C_{\alpha_{1} \dots \alpha_{n}} =
\end{equation}
\[
=  \langle\int_{\Sigma}\hspace{- 0.2cm }\; {\cal{O}}^{(2)}_{\alpha_1}\dots \int_{\Sigma}\hspace{- 0.2cm }\; {\cal{O}}_{\alpha_{n-3}}^{(2)}{\cal{O}}_{\alpha_{n-2}}({\sigma_{n-2}}) {\cal{O}}_{\alpha_{n-1}}({\sigma_{n-1}}){\cal{O}}_{\alpha_{n}}({\sigma_{n}})\rangle_0 
\]

\begin{equation}\label{c111}
 {\cal{O}}_{\alpha}^{(2)}(\sigma) = \oint_{C_{\sigma}}G^{-}\oint_{C'_{\sigma}} \bar{G}^{-}\;{\cal{O}}_{\alpha}(\sigma)
\end{equation}

It is important to state that we are not implying that these correlation function are the same as the ones containing operators ${\cal{O}}_{\alpha_1}\phi\dots {\cal{O}}_{\alpha_{n-3}}\phi\;{\cal{O}}_{\alpha_{n-2}}{\cal{O}}_{\alpha_{n-1}}{\cal{O}}_{\alpha_n}$. The above notation is simply a convenient choice for the future definition of the operator ${\cal{W}}_s$ that will act treating $G^{-},\bar{G}^{-}$ analogously to $\phi$. In any case from now on we will never use the gravitational multiplet again and we will only deal with BCOV type correlation functions. Also note that on the left hand side of (\ref{c11}) we are incorporating the integrals inside the definition of $\langle \dots\rangle_0$, while on the right they are explicit.

 ${\cal{O}}_{\alpha}^{(2)}$ is also the solution for the last step of the descent equation

\[
0 = [Q,{\cal{O}}_{\alpha}]
\]
\begin{equation}\label{c3}
d{\cal{O}}_{\alpha} = [Q,{\cal{O}}_{\alpha}^{(1)}]
\end{equation}
\[
d{\cal{O}}_{\alpha}^{(1)} = [Q,{\cal{O}}_{\alpha}^{(2)}]
\]
\[
d{\cal{O}}_{\alpha}^{(2)} = 0
\]

If ${\cal{O}}_{\alpha}$ has equal left and right U(1) charge $q^l_{\alpha} = q^r_{\alpha} = q_{\alpha} $, then ${\cal{O}}_{\alpha}^{(2)}$ has left - right charge $ q_{\alpha}-1$.

Amplitudes at genus $g\geq 2$ are defined saturating the remaining part of the Riemann surface moduli space integral by the measure $\prod_{b=1}^{3g-3}(G^{-},\mu_{b})(\bar{G}^{-},\bar{\mu}_{\bar{b} }) $, with $\mu_{b},\bar{\mu}_{\bar{b}}$ the Beltrami differentials: 

\begin{equation}\label{c1tt}
C_{i_1\dots i_n}^g =\int_{{\cal{M}}_{g}}\prod_{a=1}^{3g-3}dm^a\wedge d\bar{m}^{\bar{a}}\;\;\langle \prod_{b=1}^{3g-3}(G^{-},\mu_{b})(\bar{G}^{-},\bar{\mu}_{\bar{b} }) \int_{\Sigma_g}\hspace{- 0.2cm } \; {\cal{O}}_{\alpha_1}^{(2)}\dots\int_{\Sigma_g}\hspace{- 0.2cm } \; {\cal{O}}_{\alpha_n}^{(2)}\rangle_{g} 
\end{equation} 

The definition at genus 1 is slightly different and we will deal with it in section \ref{sez1}.

Being $d$ the total number of $G^-,\bar{G}^-$ insertions ( both around the operators and in the path integral measure ) the selection rule for the left ( and right ) U(1) R-symmetry charges $q_{\alpha_i}$ for the operator insertions dictated by the U(1) anomaly on a Calabi Yau manifold is

\begin{equation}\label{c31}
-d + \sum_{i=1}^n q_{\alpha_i} = 3 -3g  
\end{equation}

This makes an obvious parallel between (\ref{c1}) and the above formula; when the operators are marginal ( $q_{\alpha_i} = 1$ ) and $d_i=0,1$ in (\ref{c1}), the total gravitational degree $d$ equals the number of degree one gravitational descendant operators and matches the above total number of $G^-,\bar{G}^-$ insertions, again $d$. Moreover when the genus is either zero or one all the  $G^-,\bar{G}^-$ insertions come from the operators, as is the case with gravitational descendants. And it is also true that both $\phi$ and $G^-,\bar{G}^-$ serve as an integral over ${\cal{M}}_{g,n}$.

The analogy between the two cases in fact cannot be pushed further, their ghost number and behaviour under the action of $Q$ being opposite. But this will not affect our purposes.
 
It is still possible for BCOV type amplitudes on Calabi Yau to contain gravitational descendants like (\ref{b1}) but being their total ghost number $q_{\alpha}+d$ and having the selection rule (\ref{c31}), they can appear on genus $g$ amplitudes only when $d=1$ and ${\cal{O}}_{\alpha}={\cal{O}}_{0}=1$ ( $q_{\alpha} = 0$ ), the so called dilaton operator ( the case $g=0$ allows a little more ). We will not consider them in the future. 

\textbf{Remark}: the main difference between topological amplitudes of Witten and BCOV type is that in the first case the gravitational multiplet is distinct from the matter multiplets, while in the second case the supercurrents $G^{-},\bar{G}^{-}$ are constructed with the same fields entering the matter operators. Consequently the requirement for forming a volume form for the moduli space integral out of gravitational multiplet fields (\ref{c1}) and any eventual U(1) charge anomaly for the matter fields, merge in the BCOV case into the unique condition (\ref{c31}). 

From now on we differentiate the notation for the indexes of the operators depending on the U(1) charge associated. We will indicate a generic matter operator with greek indexes $\alpha,\beta,\gamma \dots$, $q=1$ ( marginal )  operators with mid alphabet latin indexes $i,j,k,\dots$, $q=2$ operators with beginning of the alphabet latin letters $a,b,c,\dots$ and the single $q=3$ and $q=0$ operators respectively with $x$  and  $0$. 

Amplitudes involving marginal operators can be obtained by appropriate derivatives 

\begin{eqnarray}\label{c32}
\langle {\cal{O}}_{1,i_{1}}\dots{\cal{O}}_{1,i_{n}}\rangle_{g(\geq 1)} &=& C^{g(\geq 1)}_{i_{1}\dots i_{n}} = D_{i_1}\dots D_{i_n}\langle 1\rangle_{g(\geq 1)} \\
\langle {\cal{O}}_{1,i_{1}}\dots{\cal{O}}_{1,i_{n-3}}{\cal{O}}_{i_{n-2}}{\cal{O}}_{i_{n-1}}{\cal{O}}_{i_{n}}\rangle_0 &=& C_{i_{1}\dots i_{n}} = D_{i_1}\dots D_{i_{n-3}}\langle  {\cal{O}}_{i_{n-2}}{\cal{O}}_{i_{n-1}}{\cal{O}}_{i_{n}}\rangle_0 \\
\langle {\cal{O}}_{i_{n-2}}{\cal{O}}_{i_{n-1}}{\cal{O}}_{i_{n}}\rangle_0 &=&  C_{i_{n-2}i_{n-1} i_n} = \partial_{i_{n-2}}\partial_{i_{n-1}} \partial_{i_n}F
\end{eqnarray}

with $F$ called  the prepotential.

Two metrics can be introduced on the moduli space, both covariantly constant under the action of $D_{i}= \partial_{i} - A_{i}$ and $\bar{D}_{\bar{i}}= \bar{\partial}_{\bar{i}} - A_{\bar{i}}$:

\begin{equation}
\eta_{\alpha\beta}, \;\;\;\;\; D_{i}\eta_{\alpha\beta} =\bar{D}_{\bar{i}}\eta_{\alpha\beta} =0
\end{equation} 

\begin{equation}
g_{\alpha\bar{\beta}}, \;\;\;\;\; D_{i}g_{\alpha\bar{\beta}} =\bar{D}_{\bar{i}}g_{\alpha\bar{\beta}} =0
\end{equation} 

The last equation in particular fixes the connections to be

\begin{equation}\label{p12}
A_{ik}^{\;l} = g^{i\bar{j}}\partial_kg_{\bar{j}l} \;\;\; A_{\bar{i}\bar{k}}^{\;\bar{l}} = g^{\bar{i}j}\bar{\partial}_{\bar{k}}g_{j\bar{l}} 
\end{equation}

and the mixed indexes components vanishing. Under complex conjugation $g$ goes to itself, $(g_{\alpha\bar{\beta}})^{*} = g_{\bar{\alpha}\beta}$, while $\eta$ transforms into the corresponding metric for the antitoplogical moduli space, $(\eta_{\alpha\beta})^{*}=\bar{\eta}_{\bar{\alpha}\bar{\beta}}$. 

We introduce the tt* equations of \cite{Cecotti:1991me}. These are equivalent to the existence of a flat connection given by the sum of the usual metric connection with the chiral ring matrix $C_i$ ( the three point function with one index raised ) and its complex conjugate $\bar{C}_{\bar{i}}$:

\begin{equation}\label{p11}
\nabla_{i} \equiv D_{i} - k^{-1} C_{i} \;\;\; \;\; \bar{\nabla}_{\bar{i}} \equiv \bar{D}_{\bar{i}} -k\bar{C}_{\bar{i}}
\end{equation}
\[
[\nabla_{i},\bar{\nabla}_{\bar{j}}] = [\bar{\nabla}_{\bar{i}},\bar{\nabla}_{\bar{j}}] = [\nabla_{i},\nabla_{j}] = 0 
\]

with $k$ a generic coefficient which will be fixed from now on to the value $\frac{1}{2}$ \footnote{the reason behind this choice will become clear soon}. 

Moreover there exist a matrix $M_{\bar{\alpha}}^{\beta}$  defined as

\begin{equation}
M_{\bar{\alpha}}^{\beta} = g_{\bar{\alpha}\gamma}\eta^{\gamma\beta}
\end{equation} 

which is by construction invariant under parallel transport by $D_{i}$ and $\bar{D}_{\bar{i}}$, and also by the flat derivatives $\nabla_{i}$ and $\bar{\nabla}_{\bar{i}}$ because

\[
M_{\bar{\alpha}}^{\rho}C_{i\rho}^{\beta} =  C_{i\bar{\alpha}}^{\beta} = C_{i\bar{\alpha}}^{\bar{\rho}}   M_{\bar{\rho}}^{\beta}
\]

and similarly for the action of $\bar{C}_{\bar{i}}$. It can be normalized as

\begin{equation}\label{norm}
M\bar{M}=1
\end{equation}

The worldsheet construction of $\eta_{\alpha\beta}$ is simply the sphere with two fixed operator insertions ${\cal{O}}_{\alpha}$ and ${\cal{O}}_{\beta}$ with $q_{\alpha} + q_{\beta}=3$. In fact due to the irrelevance of the positions and the trivial OPE of the identity with any other operator, $\eta$ can be seen as the three point sphere amplitude $C$ with one operator being the identity\footnote{the description as a sphere three point amplitude is more natural from an algebraic point of view as the resulting moduli space is 0 dimensional. However using the OPE rules it is possible to "absorb" the identity in one of the other operators; obviously the path integral description changes and it is necessary to divide by the CKV groups left free from having only two operator positions fixed}:

\[
\langle {\cal{O}}_{\alpha}{\cal{O}}_{\beta}{\cal{O}}_{0}\rangle_0  =  C_{\alpha\beta 0} = \eta_{\alpha\gamma}C_{0\beta}^{\gamma} =\eta_{\alpha\gamma}\delta_{\beta}^{\gamma} =\eta_{\alpha\beta} = \langle {\cal{O}}_{\alpha}   {\cal{O}}_{\beta}\rangle_0  
\]
 
The metric $g_{\alpha\bar{\beta}}$ is instead more complicated and it can be represented as two hemispheres, one topologically and the other antitopologically twisted, joined by an infinitely long tube interpolating between the two topological CFTs. On each hemisphere there is one operator insertion, ${\cal{O}}_{\alpha}$ and ${\cal{O}}_{\bar{\beta}}$ respectively, with the charge condition being $q_{\alpha} + q_{\bar{\beta}} = 0, \;(q_{\bar{\beta}} = -q_{\beta})$.

\section{genus 0}

We begin by consider a BCOV type amplitude as in (\ref{c11})

\[
\langle {\cal{O}}_{1,i_{1}}\dots{\cal{O}}_{1,i_{n-3}}{\cal{O}}_{i_{n-2}}{\cal{O}}_{i_{n-1}}{\cal{O}}_{i_{n}}\rangle_{0} 
\]

with the last three positions fixed. For convenience we will use ${\cal{O}}_{i_{n-1}}$ and ${\cal{O}}_{i_{n}}$ to play the role of ${\cal{O}}_{d_{n-1},i_{n-1}}$ and ${\cal{O}}_{d_n, i_{n}}$ inside (\ref{a}), while the operator corresponding to ${\cal{O}}_{d_1, i_{1}}$ will be ${\cal{O}}_{1, i_{1}}$. While applying (\ref{a}) we keep fixed or integrated the positions of the operators as they were originally in the above amplitude. The only exception is  ${\cal{O}}_{1, i_{1}}$ that, after the transformation ${\cal{O}}_{1, i_{1}}\rightarrow {\cal{O}}_{ i_{1}}$, will pass from being integrated to being kept fixed, as one would expect after having performed the moduli space integration over its cohomology class that has led to the recursion relation itself. We then define the action of the operator ${\cal{W}}$ on these genus zero amplitudes, that later on will be transformed into ${\cal{W}}_s$ , its "symmetrized" version. The goal is to formally reproduce the right hand side of the recursion relation (\ref{a}) when acting on BCOV type amplitudes, where the scalar $\phi$ will be replaced by the two supercurrents $G^-,\bar{G}^-$.  So in particular when moving from left to right in the expression below the transformation ${\cal{O}}_{1, i_{1}}\rightarrow {\cal{O}}_{ i_{1}}$ will mean that  ${\cal{O}}_{1, i_{1}}=\oint_{C_{\sigma}}G^{-}\oint_{C'_{\sigma}} \bar{G}^{-}\;{\cal{O}}_{i_1}(\sigma)\rightarrow {\cal{O}}_{i_1}$, instead of the original ${\cal{O}}_{1, i_{1}}=\phi{\cal{O}}_{ i_{1}}\rightarrow {\cal{O}}_{ i_{1}}$.

\begin{equation}\label{ccc}
{\cal{W}}\langle {\cal{O}}_{1,i_{1}}\dots{\cal{O}}_{1,i_{n-3}}{\cal{O}}_{i_{n-2}}{\cal{O}}_{i_{n-1}}{\cal{O}}_{i_{n}}\rangle_{0} = 
\end{equation}
\[
= \hspace{-0,5cm}\sum_{X\cup Y=\{2,\dots,n-3\}}\hspace{-0,5cm}\langle {\cal{O}}_{i_{1}}{\cal{O}}_{i_{n-2}}\prod_{r\in X}{\cal{O}}_{1,i_{r}}{\cal{O}}_{\alpha}\rangle_{0}\eta^{\alpha\beta}\langle {\cal{O}}_{\beta}\prod_{t\in Y}{\cal{O}}_{1,i_{t}}{\cal{O}}_{i_{n-1}}{\cal{O}}_{i_{n}}\rangle_{0} +
\]
\[
+ \langle {\cal{O}}_{i_{1}}\prod_{r\in X}{\cal{O}}_{1,i_{r}}{\cal{O}}_{\alpha}\rangle_{0}\eta^{\alpha\beta}\langle {\cal{O}}_{\beta}^{\int}\prod_{t\in Y}{\cal{O}}_{1,i_{t}}{\cal{O}}_{i_{n-2}}{\cal{O}}_{i_{n-1}}{\cal{O}}_{i_{n}}\rangle_{0}
\]
\[
+ \langle {\cal{O}}_{i_{1}}\hspace{-0,2cm}\prod_{r\in X-\{s\}}\hspace{-0,2cm}{\cal{O}}_{1,i_{r}}{\cal{O}}_{i_s}{\cal{O}}_{\alpha}\rangle_{0}\;\eta^{\alpha\beta}\langle {\cal{O}}_{1,\beta}\prod_{t\in Y}{\cal{O}}_{1,i_{t}}{\cal{O}}_{i_{n-2}}{\cal{O}}_{i_{n-1}}{\cal{O}}_{i_{n}}\rangle_{0} 
\]

We want to explain the various terms. The first two lines are the direct translation of the right hand side of (\ref{a}) with the fixed position operator ${\cal{O}}_{i_{n-2}}$ first on one side and then on the other, as we kept it outside the sets $X$ and $Y$ for later convenience. The operators ${\cal{O}}_{\alpha}$ and ${\cal{O}}_{\beta}$ have their positions naturally fixed if the number of operators whose position is fixed in the same amplitude is less or equal to two. Otherwise they are integrated, and in this case the notation becomes ${\cal{O}}_{\dots}^{\int}$.  

The second line in fact immediately simplifies because

\begin{equation}\label{yyu}
\langle {\cal{O}}_{i_{1}}\prod_{r\in X}{\cal{O}}_{1,i_{r}}{\cal{O}}_{\alpha}\rangle_{0} = \langle {\cal{O}}_{i_{1}}\prod_{r\in X}{\cal{O}}_{1,i_{r}}{\cal{O}}_{a}\rangle_{0} =\prod_{r\in X}D_{i_{r}}\langle {\cal{O}}_{i_{1}}{\cal{O}}_{a}\rangle_{0} = \prod_{r\in X}D_{i_{r}}\eta_{i_1 a} = \delta_{card(X),0}
\end{equation}

where we have used first the U(1) charge condition (\ref{c31}) fixing $q_{\alpha} = 2 = q_a$, and then the fact that $\eta$ is covariantly constant. Thus contributions come only when $X=\{0\}$ and $Y$ the whole set  $ \{2,\dots,n-3\}$.

\begin{figure}[htb]
  \centering
  \vspace{-10pt}
  \def\svgwidth{470pt}
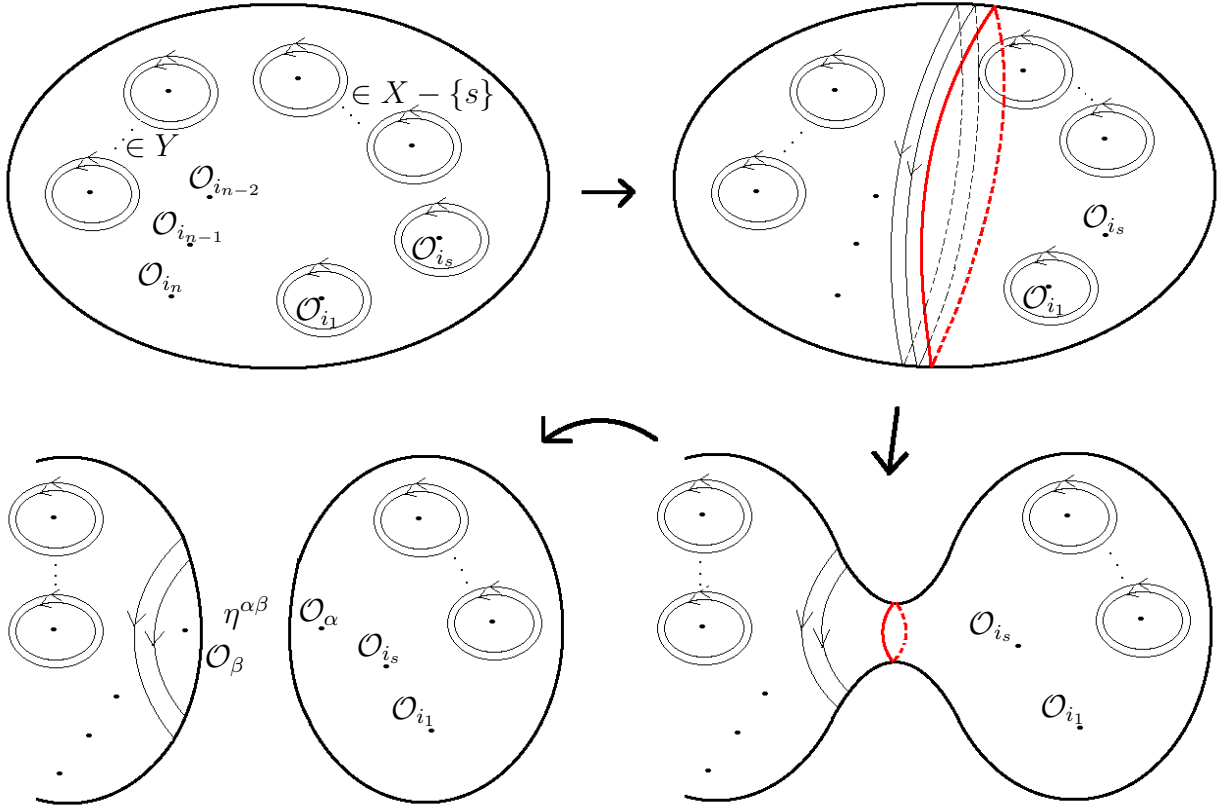
  \vspace{-50pt}  
  \caption{The correlation function on the sphere $\langle {\cal{O}}_{1,i_{1}}\dots{\cal{O}}_{1,i_{n-3}}{\cal{O}}_{i_{n-2}}{\cal{O}}_{i_{n-1}}{\cal{O}}_{i_{n}}\rangle_{0} $  is considered. After the first arrow we deform the contour of the two supercurrents $G^-,\bar{G}^-$  of ${\cal{O}}_{1,i_s}$ to surround also the operators  ${\cal{O}}_{1,i_{1}}\prod_{r\in X-\{s\}}{\cal{O}}_{1,i_{r}}$. One possible cycle which is going to shrink to a node is depicted in red. After the second arrow we start forming the node shrinking the red cycle, and consequently ${\cal{O}}_{1,i_1}\rightarrow{\cal{O}}_{i_1}$.  In this phase the position of the operator ${\cal{O}}_{i_{s}}$ is fixed with respect to the position of the shrinking cycle, while the latter is integrated. When the node is replaced by ${\cal{O}}_{\alpha}\eta^{\alpha\beta}{\cal{O}}_{\beta}$ the two supercurrents will encircle the operator ${\cal{O}}_{\beta}$ and we have the situation following the third arrow.}
\label{figura1} 
\end{figure}

The last term is due to a subtle but important point we have to consider when we apply the machine leading to (\ref{a})  to amplitudes of BCOV type. In particular the passage from the left to the right hand side of (\ref{a})  is achieved through the appearance of a node  as we degenerate the genus zero correlation function into two spheres sharing a single point. When we represent operators with gravitational degree one as (\ref{c111}), it is possible for the node to arise in between the position of one of the operators and the two circle integrals of the supercurrents $G^-, \bar{G}^- $; more specifically if we think of the node as appearing after the shrinking to a point of a nontrivial cycle on the Riemann surface, this cycle can be taken as well in an intermediate worldsheet time between the one of the operator and its two accompanying supercurrents. Furthermore it is possible to include inside the region bounded by the shrinking cycle other  operators as well, as long as they are closed under the action of $G^-, \bar{G}^- $. This construction is represented in figure (\ref{figura1}).

The net result is that the two supercurrents $G^-, \bar{G}^- $ originally encircling around one operator, for example $O_{i_s}$, after the recursion relation has been applied will surround the position of the node on the opposite sphere. Thus the gravitational coupling shifts from $O_{1,i_s}$ to the operator $O_{\beta}^{\int}$ replacing the node on that side, and transforming it to $O_{1,\beta}$. Note that the integrated position of the operator $O_{1,i_s}$ on the original amplitude becomes fixed on the factorized sphere, because the integration passes from the position of  $O_{1,i_s}$ to the position of the node as should be clear from the picture.

Having defined the action of ${\cal{W}}$ let us step aside for a moment and discuss a point we have so far avoided. It is well known that BCOV type amplitudes possess a nontrivial dependence on antiholomorphic moduli, while it is their holomorphic limit that resembles the Witten's definition for two dimensional   topological gravity + matter theories leading to the recursion relations. Thus we need to supplement  equation (\ref{ccc}) with the correct antiholomorphic moduli dependence. In fact this is a necessity as the right hand side of (\ref{ccc}) identically vanishes if we apply the U(1) charge condition (\ref{c31}). For this reason let us consider the following algebra:

\begin{equation}\label{cr1}
[{\cal{W}}_s,\bar{\nabla}_{\bar{i}}] = \bar{\nabla}_{\bar{i}} 
\end{equation}
\begin{equation}\label{cr2}
[{\cal{W}}_s,\nabla_{i}] = 0
\end{equation}
\begin{equation}\label{cr3}
[\nabla_{i},\bar{\nabla}_{\bar{j}}] = [\nabla_{i},\nabla_j] = [\bar{\nabla}_{\bar{i}},\bar{\nabla}_{\bar{j}}]=0
\end{equation}

The main point of the paper will be to show that the first commutator is equivalent to the H.A.E. . The second commutator will be discussed later on and  we have already met the last.  

So we want to evaluate (\ref{cr1}) when acting on genus zero amplitudes with $n$ marginal operator insertions. The right hand side is easy. We recall the definition $\bar{\nabla}_{\bar{i}} \equiv \bar{\partial}_{\bar{i}} - A_{\bar{i}} - \frac{1}{2}\bar{C}_{\bar{i}}$ and, because the correlation function contains only holomorphic indexes, $A_{\bar{i}}$ vanishes, as it would be a mixed component. Moreover $\bar{C}_{\bar{i}}$ acting on low indexes transforms charge $q$ into charge $q-1$ operators, thus killing any amplitude previously satisfying the U(1) charge selection rule. Thus what survives is only the antiholomorphic simple derivative $ \bar{\nabla}_{\bar{i}}C_{i_1\dots i_n} = \bar{\partial}_{\bar{i}}C_{i_1\dots i_n}  $. 

The left  hand side is more tricky: $-\bar{\nabla}_{\bar{i}}\left({\cal{W}}_sC_{i_1\dots i_n}\right)$ is well defined, as it is simply the action of $\bar{\nabla}_{\bar{i}}$ on the right hand side of (\ref{ccc});  instead we should explain what we mean for ${\cal{W}}_s\left(\bar{\nabla}_{\bar{i}}C_{i_1\dots i_n}\right)$. We have defined the action of ${\cal{W}}$ on amplitudes with topological marginal operators, now we associate to $\bar{\nabla}_{\bar{i}}$ an operator insertion ${\cal{O}}_{\bar{\nabla}_{\bar{i}}}$ and we extend the definition in the natural way.

${\cal{O}}_{\bar{\nabla}_{\bar{i}}}$ is a short  distance refinement of the operator associated to $\bar{D}_{\bar{i}}$ , which is  the integral of $\oint_{C_{\sigma}}G^{+}\oint_{C'_{\sigma}}\bar{G}^{+} \bar{{\cal{O}}}_{\bar{i}}(\sigma)$ . The difference between $\bar{\nabla}_{\bar{i}}$ and $\bar{D}_{\bar{i}}$ is the action of $-\frac{1}{2}\bar{C}_{\bar{i}}$ on the indexes of the operators inside the amplitude, that is $-\frac{1}{2}$ the coefficient of the OPE of the operators with $\bar{{\cal{O}}}_{\bar{i}}$ \footnote{there is also a nontrivial action on the vacuum bundle ${\cal{L}}^{2-2g-n}$ to which the amplitude belongs, but for antiholomorphic derivatives and topological amplitudes it vanishes.}. In particular if the index belongs to an operator $ {\cal{O}}_{1,\alpha}$ the OPE should be taken under the action of the two spin two supercurrents $G^{-}$ and $\bar{G}^{-}$  that is \footnote{ note the minus sign when $\bar{C}_{\bar{i}}$ acts on low indexes in accordance with the usual convention for a connection} 

\[
-\frac{1}{2}\bar{C}_{\bar{i}}\langle \dots {\cal{O}}_{1,\alpha} \dots \rangle = \frac{1}{2}\bar{C}_{\bar{i}\alpha}^{\beta}\langle \dots \int_{\Sigma}\oint_{C_{\sigma}}G^{-}\oint_{C'_{\sigma}}\bar{G}^{-} {\cal{O}}_{\beta}(\sigma)\dots \rangle + \dots = \frac{1}{2}\bar{C}_{\bar{i}\alpha}^{\beta}\langle \dots {\cal{O}}_{1,\beta} \dots \rangle + \dots
\]

with $q_{\beta} = q_{\alpha}-1$. Thus  ${\cal{O}}_{\bar{\nabla}_{\bar{i}}}$ is defined as the sum of the operator insertion corresponding to $\bar{D}_{\bar{i}}$ plus the insertion of $\frac{1}{2}\bar{{\cal{O}}}_{\bar{i}}$ inside an infinitesimal neighborhood $\Delta_u$ around the position of each preexisting operator, reproducing the OPEs of $-\frac{1}{2}\bar{C}_{\bar{i}}$. Moreover we include the short distance regularization prescription for $\oint_{C_{\sigma}}G^{+}\oint_{C'_{\sigma}}\bar{G}^{+} \bar{{\cal{O}}}_{\bar{i}}(\sigma)$ when approaching other operators, see \cite{Bershadsky:1993cx}, that can be translated as excluding from its region of integration a small neighborhood around each operator:

\begin{equation}\label{tra}
{\cal{O}}_{\bar{\nabla}_{\bar{i}}} \equiv \int_{\Sigma-\{\Delta_u\}}\;\oint_{C_{\sigma}}G^{+}\oint_{C'_{\sigma}}\bar{G}^{+} \bar{{\cal{O}}}_{\bar{i}}(\sigma) + \frac{1}{2}\sum_{u=1}^n\int_{\Delta_{u}}\bar{{\cal{O}}}_{\bar{i}} \;\;\footnote{this definition for ${\cal{O}}_{\bar{\nabla}_{\bar{i}}} $ is clearly dependent on the choice of $\Delta_{u}$. However we will use it only as an intermediate step before going back to the matrix $\bar{C}_{\bar{i}}$, thus $\Delta_{u} $ will drop out in the final result.}
\end{equation}

The action of ${\cal{W}}$ on the amplitude containing ${\cal{O}}_{\bar{\nabla}_{\bar{i}}}$ is the natural generalization of (\ref{ccc}), with ${\cal{O}}_{\bar{\nabla}_{\bar{i}}}$ treated similarly to the other operators but conveniently kept outside the sets $X$ and $Y$ and thus appearing first on one side and then on the other.

\begin{equation}\label{ccc2}
{\cal{W}}\langle {\cal{O}}_{1,i_{1}}\dots{\cal{O}}_{1,i_{n-3}}{\cal{O}}_{i_{n-2}}{\cal{O}}_{i_{n-1}}{\cal{O}}_{i_{n}}{\cal{O}}_{\bar{\nabla}_{\bar{i}}}\rangle_{0} = 
\end{equation}
\[
= \hspace{-0,5cm}\sum_{X\cup Y=\{2,\dots,n-3\}}\hspace{-0,5cm}\langle {\cal{O}}_{i_{1}}{\cal{O}}_{i_{n-2}}\prod_{r\in X}{\cal{O}}_{1,i_{r}}{\cal{O}}_{\bar{\nabla}_{\bar{i}}}{\cal{O}}_{\alpha}\rangle_{0}\eta^{\alpha\beta}\langle {\cal{O}}_{\beta}\prod_{t\in Y}{\cal{O}}_{1,i_{t}}{\cal{O}}_{i_{n-1}}{\cal{O}}_{i_{n}}\rangle_{0} +
\]
\[
 +\langle {\cal{O}}_{i_{1}}{\cal{O}}_{i_{n-2}}\prod_{r\in X}{\cal{O}}_{1,i_{r}}{\cal{O}}_{\alpha}\rangle_{0}\eta^{\alpha\beta}\langle {\cal{O}}_{\beta}\prod_{t\in Y}{\cal{O}}_{1,i_{t}}{\cal{O}}_{i_{n-1}}{\cal{O}}_{i_{n}}{\cal{O}}_{\bar{\nabla}_{\bar{i}}}\rangle_{0} +
\]
\[
+\langle {\cal{O}}_{i_{1}}\hspace{-0,2cm}\prod_{r\in X-\{s\}}\hspace{-0,2cm}{\cal{O}}_{1,i_{r}}{\cal{O}}_{i_s}\bar{{\cal{O}}}_{\bar{\nabla}_{\bar{i}}}{\cal{O}}_{\alpha}\rangle_{0}\;\eta^{\alpha\beta}\langle {\cal{O}}_{1,\beta}\prod_{t\in Y}{\cal{O}}_{1,i_{t}}{\cal{O}}_{i_{n-2}}{\cal{O}}_{i_{n-1}}{\cal{O}}_{i_{n}}\rangle_{0} 
\]
\[
+ \langle {\cal{O}}_{i_{1}}\hspace{-0,2cm}\prod_{r\in X-\{s\}}\hspace{-0,2cm}{\cal{O}}_{1,i_{r}}{\cal{O}}_{i_s}{\cal{O}}_{\alpha}\rangle_{0}\;\eta^{\alpha\beta}\langle {\cal{O}}_{1,\beta}\prod_{t\in Y}{\cal{O}}_{1,i_{t}}{\cal{O}}_{i_{n-2}}{\cal{O}}_{i_{n-1}}{\cal{O}}_{i_{n}}{\cal{O}}_{\bar{\nabla}_{\bar{i}}}\rangle_{0} 
\]
\[
+ \langle {\cal{O}}_{i_{1}}\bar{{\cal{O}}}_{\bar{\nabla}_{\bar{i}}}{\cal{O}}_{\alpha}\rangle_{0}\eta^{\alpha\beta}\langle {\cal{O}}_{\beta}^{\int}{\cal{O}}_{1,i_{2}}\dots{\cal{O}}_{1,i_{n-3}} {\cal{O}}_{i_{n-2}}{\cal{O}}_{i_{n-1}}{\cal{O}}_{i_{n}}\rangle_{0}
\]
\[
+ \langle {\cal{O}}_{i_{1}}{\cal{O}}_{\alpha}\rangle_{0}\eta^{\alpha\beta}\langle {\cal{O}}_{\beta}^{\int}{\cal{O}}_{1,i_{2}}\dots{\cal{O}}_{1,i_{n-3}} {\cal{O}}_{i_{n-2}}{\cal{O}}_{i_{n-1}}{\cal{O}}_{i_{n}}{\cal{O}}_{\bar{\nabla}_{\bar{i}}}\rangle_{0}
\]

Written in this way the definition is apparently inconsistent because of the third term on the right hand side. This has been generated through the mechanism discussed previously, when the Riemann surface node makes its appearance between one operator ${\cal{O}}_{i_s}$ and its two accompanying supercurrents. However the requirement was that, after the Riemann surface degeneration has been applied, all the operators present on the same sphere containing ${\cal{O}}_{i_s}$ have to be closed under the action of $G^-, \bar{G}^- $. And this fails for $\oint_{C_{\sigma}}G^{+}\oint_{C'_{\sigma}}\bar{G}^{+} \bar{{\cal{O}}}_{\bar{i}}(\sigma)$ in the definition (\ref{tra}) of ${\cal{O}}_{\bar{\nabla}_{\bar{i}}}$. This object should also disappear from the fifth term as a consequence of $\bar{D}_{\bar{i}}\eta = 0$. However there is a more fundamental overall reason why the first term in (\ref{tra}) cannot appear anywhere in the above expression, and it is again the U(1) charge selection rule. For example the first term requires $q_{\alpha} + 2 + q_{\bar{\nabla}_{\bar{i}}} = 3$ for the first sphere, $q_{\alpha} + q_{\beta} = 3$ for the metric and $q_{\beta} + 2  = 3$ for the second sphere, with $q_{\bar{\nabla}_{\bar{i}}}$ either zero or $-1$ depending on which term inside ${\cal{O}}_{\bar{\nabla}_{\bar{i}}} $ you are considering. Obviously the three conditions are satisfied together only when $q_{\bar{\nabla}_{\bar{i}}} =-1 $; this can be straightforwardly derived for every term. Thus the result is that the only non vanishing contributions select  the $\frac{1}{2}\bar{{\cal{O}}}_{\bar{i}}$ piece from ${\cal{O}}_{\bar{\nabla}_{\bar{i}}}$.  

It is here crucial the interpretation of the action of $\bar{\nabla}_{\bar{i}}$ as the operator insertion of ${\cal{O}}_{\bar{\nabla}_{\bar{i}}} $. Once ${\cal{W}}$ has been applied the insertion of ${\cal{O}}_{\alpha}$ or ${\cal{O}}_{\beta}$ are defined to be consistent with the definition (\ref{tra}), and this in particular means that the OPE with $\frac{1}{2}\bar{{\cal{O}}}_{\bar{i}}$ ( represented by the second term in (\ref{tra}) ) includes also the action on ${\cal{O}}_{\alpha}$ or ${\cal{O}}_{\beta}$, together with every other operator in the amplitude.

Understood this point we can finally go back from the  ${\cal{O}}_{\bar{\nabla}_{\bar{i}}} $ - operatorial description to the  to $\bar{\nabla}_{\bar{i}}$ - covariant derivative formalism, and this translates into the selection of the $-\frac{1}{2}\bar{C}_{\bar{i}}$ term inside $\bar{\nabla}_{\bar{i}}$ acting on every index inside the two factorized amplitudes, with $\bar{D}_{\bar{i}}$ decoupling.

The evaluation of $\bar{\nabla}_{\bar{i}}\left({\cal{W}}C_{i_1\dots i_n}\right)$ is easier. This object consists of the insertion of the integral of $\oint G^{+}\oint\bar{G}^{+} \bar{{\cal{O}}}_{\bar{i}}$ inside all the amplitudes and the metric, plus the action of $-\frac{1}{2}\bar{C}_{\bar{i}}$ on every low and high index. Again the U(1) charge condition selects only the latter.  The difference between ${\cal{W}}\left(\bar{\nabla}_{\bar{i}}C_{i_1\dots i_n}\right)$  and $\bar{\nabla}_{\bar{i}}\left({\cal{W}}C_{i_1\dots i_n}\right)$ then reduces to minus the action of $-\frac{1}{2}\bar{C}_{\bar{i}}$ on the metric $\eta^{\alpha\beta}$. Taking care of the correct $q_{\alpha}$ and $q_{\beta}$ charges it gives

\begin{equation}\label{pt}
{\cal{W}}\left(\bar{\nabla}_{\bar{i}}C_{i_1\dots i_n}\right) - \bar{\nabla}_{\bar{i}}\left({\cal{W}}C_{i_1\dots i_n}\right) = 
\end{equation}
\[
= \sum_{X\cup Y=\{2,\dots,n-3\}}\hspace{-0,5cm}\langle {\cal{O}}_{i_{1}}{\cal{O}}_{i_{n-2}}\prod_{r\in X}{\cal{O}}_{1,i_{r}}{\cal{O}}_{l}\rangle_{0}\;\frac{1}{2}\left(\bar{C}_{\bar{i}a}^{l}\eta^{am} + \eta^{lb}\bar{C}_{\bar{i}b}^{m} \right)\langle {\cal{O}}_{m}\prod_{t\in Y}{\cal{O}}_{1,i_{t}}{\cal{O}}_{i_{n-1}}{\cal{O}}_{i_{n}}\rangle_{0} +
\]
\[
+\langle {\cal{O}}_{i_{1}}\hspace{-0,2cm}\prod_{r\in X-\{s\}}\hspace{-0,2cm}{\cal{O}}_{1,i_{r}}{\cal{O}}_{i_s}{\cal{O}}_{l}\rangle_{0}\;\frac{1}{2}\left(\bar{C}_{\bar{i}a}^l\eta^{am} + \eta^{lb}\bar{C}_{\bar{i}b}^m\right)\langle {\cal{O}}_{1,m}\prod_{t\in Y}{\cal{O}}_{1,i_{t}}{\cal{O}}_{i_{n-2}}{\cal{O}}_{i_{n-1}}{\cal{O}}_{i_{n}}\rangle_{0} +
\]
\[
+\langle {\cal{O}}_{i_{1}}{\cal{O}}_{a}\rangle_{0}\frac{1}{2}\left( \bar{C}_{\bar{i}x}^{a}\eta^{x0} + \eta^{am}\bar{C}_{\bar{i}m}^{0} \right)\langle {\cal{O}}_{0}^{\int}{\cal{O}}_{1,i_{2}}\dots{\cal{O}}_{1,i_{n-3}} {\cal{O}}_{i_{n-2}}{\cal{O}}_{i_{n-1}}{\cal{O}}_{i_{n}}\rangle_{0} 
\]

Obviously $\frac{1}{2}\bar{C}_{\bar{i}a}^{l}\eta^{am} + \frac{1}{2}\eta^{lb}\bar{C}_{\bar{i}b}^{m} =\bar{C}_{\bar{i}}^{lm} $.  Moreover, remembering the worldsheet definition of $\eta$, $\langle {\cal{O}}_{i_{1}}{\cal{O}}_{a}\rangle_{0}=\eta_{i_1 a}$ , and the normalization (\ref{norm}), $M\bar{M}=1$, we have

\[\eta_{i_1 a}\bar{C}_{\bar{i}x}^{a}\eta^{x0} = \eta_{i_1 a}g^{a\bar{a}}\bar{C}_{\bar{i}\bar{a}}^{0} = \eta_{i_1 a}g^{a\bar{a}}\bar{C}_{\bar{i}\bar{a}\bar{0}}g^{0\bar{0}} = \eta_{i_1 a}g^{a\bar{a}}\bar{\eta}_{\bar{a}\bar{i}}g^{0\bar{0}} = g_{i_1\bar{i}}g^{0\bar{0}} \equiv G_{i_1\bar{i}}
\]
\[\eta_{i_1 a}\eta^{am} \bar{C}_{\bar{i}m}^{0}= \bar{C}_{\bar{i}i_1}^{0} = \eta_{i_1b}\bar{C}_{\bar{i}\bar{0}}^{b}g^{0\bar{0}} = \eta_{i_1b}g^{b\bar{b}}\bar{C}_{\bar{i}\bar{b}\bar{0}}g^{0\bar{0}} =  \eta_{i_1b}g^{b\bar{b}}\bar{\eta}_{\bar{b}\bar{i}}g^{0\bar{0}} = g_{i_1\bar{i}}g^{0\bar{0}} \equiv G_{i_1\bar{i}}
\]

Finally let consider the amplitude containing ${\cal{O}}_{0}^{\int}=\int_{\Sigma}{\cal{O}}_{0}$ ( in this case $\Sigma = S^2$ ). The operator ${\cal{O}}_{0}$ has been previously identified with the identity 1 but this is not really valid globally on the Riemann surface. The correct prescription is ${\cal{O}}_{0} = e^{R/2\pi}$ with $R$ the  Riemann surface curvature. The rationale behind this identification is that in a topologically twisted theory of BCOV type the action is modified with the addition of a term implementing the topological twist:

\[
\frac{1}{2}\int_{\Sigma} R\varphi 
\]

with $\varphi$ the scalar bosonizing the U(1) R-symmetry current. Due to conformal invariance $R$ can be represented as a sum of $2g-2+n$ delta functions on a surface of genus $g$ with $n$ operator insertions, each delta function carrying $\pm 2\pi$ units of curvature. Thus the above integral reduces to an insertion of $e^{\varphi/2}$ at each of these $2g-2+n$ points. On the other side it is known that there exists a contact term between the operator ${\cal{O}}_{0}$ and any other marginal operator ${\cal{O}}_{i}^{(2)}$: this was computed in \cite{Bershadsky:1993cx} to be $-\partial_i K \frac{R}{2\pi} $, $K$ the moduli space Kahler potential, and explained as arising from an hidden contact term between these operator insertions ${\cal{O}}_{i}^{(2)}$ and $\frac{1}{2}\int_{\Sigma} R\varphi$ inside the action, that we have seen corresponding $2g-2+n$ insertions of $e^{\varphi/2}$. Thus ${\cal{O}}_{0}$ is defined in such a way as to reduce locally to the identity, but on the points where $e^{\varphi/2}$ is inserted in which case the short distance divergent coefficient with ${\cal{O}}_{i}^{(2)}$ has to be ${\cal{O}}_{i}^{(2)}$ itself at that point, thus producing the OPE with $e^{\varphi/2}$ as the final result. And this leads exactly to ${\cal{O}}_{0} = e^{R/2\pi}$.

If the operator ${\cal{O}}_{0}$ is at fixed position we can choose the delta function supports to be away 
and simply write ${\cal{O}}_{0}=1$. But if the operator is integrated the two dimensional integral selects the two form component of ${\cal{O}}_{0} = e^{R/2\pi}$ and we have $\int_{\Sigma}{\cal{O}}_{0}=\int_{\Sigma}e^{R/2\pi}=\int_{\Sigma}R/2\pi=2-2g-n$. This is indeed reminiscent of the dilaton equation appearing in \cite{Verlinde:1990ku}, identifying the dilaton ${\cal{O}}_0\phi = 1\cdot\phi$ with $-{\cal{O}}_{0}^{\int}$. Indeed we would expect that $\phi$ translates to the circle integral of the two spin two supercurrents, as for the other operators. However it is clear that such result is zero as the OPE of the identity with $G^-$ and $\bar{G}^-$ is trivial. The next possibility is the integral of the identity without the two supercurrents, giving ${\cal{O}}_{0}^{\int}$. In our case this allows us to get rid of ${\cal{O}}_{0}^{\int}$ substituting it with the factor $2-(n-1)=3-n$. 

Collecting all the terms we arrive at the result

\begin{equation}\label{ptent}
[{\cal{W}},\bar{\nabla}_{\bar{i}}]\langle {\cal{O}}_{1,i_{1}}\dots{\cal{O}}_{1,i_{n-3}}{\cal{O}}_{i_{n-2}}{\cal{O}}_{i_{n-1}}{\cal{O}}_{i_{n}}\rangle_{0} = 
\end{equation}
\[
= \sum_{X\cup Y=\{2,\dots,n-3\}}\hspace{-0,5cm}\langle {\cal{O}}_{i_{1}}{\cal{O}}_{i_{n-2}}\prod_{r\in X}{\cal{O}}_{1,i_{r}}{\cal{O}}_{l}\rangle_{0}\;\bar{C}_{\bar{i}}^{lm}\langle {\cal{O}}_{m}\prod_{t\in Y}{\cal{O}}_{1,i_{t}}{\cal{O}}_{i_{n-1}}{\cal{O}}_{i_{n}}\rangle_{0} + 
\]
\[
+ \sum_{\{2,\dots,n-3\}=X\cup Y}\hspace{-0,5cm}\langle {\cal{O}}_{i_{1}}\hspace{-0,2cm}\prod_{r\in X-\{s\}}\hspace{-0,2cm}{\cal{O}}_{1,i_{r}}{\cal{O}}_{i_s}{\cal{O}}_{l}\rangle_{0}\;\bar{C}_{\bar{i}}^{lm}\langle {\cal{O}}_{1,m}\prod_{t\in Y}{\cal{O}}_{1,i_{t}}{\cal{O}}_{i_{n-2}}{\cal{O}}_{i_{n-1}}{\cal{O}}_{i_{n}}\rangle_{0} -
\]
\[
-(n-3)G_{i_1\bar{i}}\langle {\cal{O}}_{1,i_{2}}\dots{\cal{O}}_{1,i_{n-3}}{\cal{O}}_{i_{n-2}}{\cal{O}}_{i_{n-1}}{\cal{O}}_{i_{n}}\rangle_{0}
\]

This looks already similar to the right hand side of the H.A.E. at genus zero of \cite{Bershadsky:1993cx} but with an important difference in structure: here the operator ${\cal{O}}_{1,i_{1}}$ and the couple ${\cal{O}}_{i_{n-1}}{\cal{O}}_{i_{n}}$ are singled out from the very beginning and do not enter the summation spreading the remaining operators on the two degenerate spheres. Instead  the corresponding H.A.E. treats all the operators in an equivalent way and neither ${\cal{O}}_{1,i_{1}}$ nor ${\cal{O}}_{i_{n-1}}$ and ${\cal{O}}_{i_{n}}$ are fixed by construction to be on separate amplitudes. To cure this problem we need to implement the "symmetrization" of ${\cal{W}}$ into ${\cal{W}}_s$.

 The symmetrized recursion relations corresponding to  the action of ${\cal{W}}_s$ is defined by first summing over all the choices of the three selected operators indicated from now on as ${\cal{O}}_{1,i_{p}}$ ( formerly ${\cal{O}}_{1,i_{1}}$ ), and the couple ${\cal{O}}_{i_{q_1}},{\cal{O}}_{i_{q_2}}$ ( before ${\cal{O}}_{i_{n-1}}{\cal{O}}_{i_{n}}$ ), and after the recursion relations have been applied dividing by the number of possible choices inside the two factorized amplitudes. Moreover we always select the three operators at fixed positions in the starting amplitude to contain the couple ${\cal{O}}_{i_{q_1}},{\cal{O}}_{i_{q_2}}$ plus a third random operator ${\cal{O}}_{i_{q}}$ ( where the only requirement for $q$ is to not coincide with any of the indexes $\{q_1,q_2,p\}$ ). Obviously ${\cal{O}}_{1,i_{p}}$ has gravitational degree $1$, otherwise it would be impossible to apply ${\cal{W}}$ as we have defined it. Two remarks are in order. 
 
First: summing over all the choices involves $n(n-1)(n-2)/2$ summands to the sum ( as the distinction between ${\cal{O}}_{i_{q_1}},{\cal{O}}_{i_{q_2}}$ is irrelevant ), but dividing after having applied the recursion relations only reduces the number by $2x(n-x)(n-x-1)/2$, where $x$ is the number of possible choices of  $i_{p}$ inside the first factorised amplitude, when $[\#$ of elements in $X] = x-2$. The choices of the set $\{q_1,q_2\}$ inside the other amplitude give instead $(n-x)(n-x-1)/2$, and the additional factor of two is from the possibility of interchanging the two amplitudes containing respectively  the couple of operators ${\cal{O}}_{i_{q_1}},{\cal{O}}_{i_{q_2}}$ and ${\cal{O}}_{1,i_{p}}$ ( and accordingly moving two supercurrents from one side to the other for fulfilling the U(1) charge selection rules as depicted in figure (\ref{figura1}) ). Thus this procedure really increases the number of terms in the recursion relation.

Second: formally which three operators you pick up in computing the recursion relation is irrelevant; even if every choice produces results that look different, they are really the same by construction. Thus, together with the rescaling by  $x(n-x)(n-x-1)$, the whole procedure amounts only to an overall different normalization factors in front of the right hand side of (\ref{a}). This will drastically change for genus higher than 2. However keeping explicit different choices for the three operators in all the terms gives at the end a result which is manifestly symmetric in all the operators inside the amplitude.

This definition for ${\cal{W}}_s$ thus leads to the expression:

\begin{equation}\label{pcorr}
[{\cal{W}}_s,\bar{\nabla}_{\bar{i}}]C_{i_1\dots i_n} = 
\end{equation}
\[
= \hspace{-0,5cm}\sum_{p, \{q_1, q_2\} }\sum_{X\cup Y}\frac{1}{2}\bar{C}_{\bar{i}}^{lm}\langle {\cal{O}}_{i_{p}}{\cal{O}}_{i_{q}}\prod_{r\in X}{\cal{O}}_{1,i_{r}}{\cal{O}}_{l}\rangle_{0}\langle {\cal{O}}_{m}\prod_{t\in Y}{\cal{O}}_{1,i_{t}}{\cal{O}}_{i_{q_1}}{\cal{O}}_{i_{q_2}}\rangle_{0}\frac{2}{(x)(n-x)(n-x-1)} + 
 \]
\[
+\frac{1}{2}\bar{C}_{\bar{i}}^{lm}\langle {\cal{O}}_{i_{p}}\hspace{-0,2cm}\prod_{r\in X-\{s\}}\hspace{-0,2cm}{\cal{O}}_{1,i_{r}}{\cal{O}}_{i_s}{\cal{O}}_{l}\rangle_{0}\langle {\cal{O}}_{1,m}\prod_{t\in Y}{\cal{O}}_{1,i_{t}}{\cal{O}}_{i_{q}}{\cal{O}}_{i_{q_1}}{\cal{O}}_{i_{q_2}}\rangle_{0}\frac{2}{(x-1)(n-x+1)(n-x)} +
\]
\[
-(n-3)G_{p\bar{i}}\langle {\cal{O}}_{1,i_{1}}\dots\hat{p}\dots{\cal{O}}_{1,i_{n}}\rangle_{0}\frac{2}{(n-1)(n-2)} 
\]

with $X\cup Y=\{1,\dots,n\}  -\{p,q_1,q_2,q\}$.  In the last term we can immediately simplify the sum over the sets $\{q_1,q_2\}$  with the factor $2/(n-1)(n-2)$. In the other two terms instead we can abandon this heavy notation and enlarge the sets $X$ and $Y$ to $X\cup Y=\{1,n\}$, thus including the element $\{q\}$ and the sums over $p,\{q_1,q_2\}$ together with the normalization factors that correctly mod out the overcounting of the number of equivalent sets $X$ and $Y$.

The final expression of the commutator (\ref{cr1}) applied to $C_{i_1\dots i_n}$ is:

\begin{equation}\label{pcorr2}
\bar{\partial}_{\bar{i}}C_{i_1\dots i_n} =[{\cal{W}}_s,\bar{\nabla}_{\bar{i}}]C_{i_1\dots i_n}= \frac{1}{2}\bar{C}_{\bar{i}}^{lm}\hspace{-0.3 cm}\sum_{X\cup Y=\{1,n\}}C_{\prod_{r\in X}i_rm}C_{\prod_{s\in Y}i_sl} - (n-3)\sum_{p=1}^n G_{p\bar{i}}C_{i_1\dots\hat{p}\dots i_n}
\end{equation}

which is the H.A.E. of \cite{Bershadsky:1993cx} at genus zero.

\section{genus 1}\label{sez1}

We move now to genus one. We begin by rescaling the amplitude as $\langle \dots  \rangle_{1}^{New} = 6\langle\dots \rangle_{1}^{Old}$  ( where $\langle\dots \rangle_{1}^{Old}$ is the correlation function entering (\ref{b}) ), and we are led to the definition:

\begin{equation}\label{l}
{\cal{W}}\langle {\cal{O}}_{1,i_1}\dots  {\cal{O}}_{1,i_n}\rangle_{1}^{New} = \frac{1}{2}\eta^{\alpha\beta}\langle {\cal{O}}_{i_1} {\cal{O}}_{1,i_2}\dots  {\cal{O}}_{1,i_n}{\cal{O}}_{\alpha}{\cal{O}}_{\beta}\rangle_{0} + 
\end{equation}
\[
+\langle {\cal{O}}_{i_1}{\cal{O}}_{\alpha}\rangle_{0}\eta^{\alpha\beta}\langle {\cal{O}}_{\beta}^{\int}{\cal{O}}_{1,i_2}\dots{\cal{O}}_{1,i_n}\rangle_{1}^{New} + \hspace{-0.3 cm}\sum_{X\cup Y=\{2,n\}}\hspace{-0.3 cm}\langle {\cal{O}}_{i_1}\hspace{-0.3 cm}\prod_{r\in X-\{s\}}\hspace{-0.3 cm}{\cal{O}}_{1,i_r}{\cal{O}}_{i_s}{\cal{O}}_{\alpha}\rangle_{0}\eta^{\alpha\beta}\langle {\cal{O}}_{1,\beta}\prod_{t\in Y}{\cal{O}}_{1,i_t}\rangle_{1}^{New}  
\]

For simplicity of notation from now on we will drop the label "$New$".
The first term is the direct translation of the first term in (\ref{b}) with the new normalization, while the second and third ones correspond to the second term of (\ref{b}) after using the fact that the metric is covariantly constant and having again applied the mechanism depicted in figure (\ref{figura1}). Let us evaluate the commutator $[{\cal{W}}_s,\bar{\nabla}_{\bar{i}}]$. Proceeding exactly as in the genus zero case, the U(1) charge anomaly condition selects the only non zero terms to be the ones produced by the action of the matrix $\bar{C}_{\bar{i}}$ on the metric $\eta^{\alpha\beta}$.  Explicitly

\begin{equation}\label{n}
\frac{1}{4}\left( \bar{C}_{\bar{i}b}^l\eta^{bm}+ \eta^{la}\bar{C}_{\bar{i}a}^m\right)\langle {\cal{O}}_{i_1} {\cal{O}}_{1,i_2}\dots  {\cal{O}}_{1,i_n}{\cal{O}}_{l}{\cal{O}}_{m}\rangle_{0} +
\end{equation}
\[
+\frac{1}{2}\left( \bar{C}_{\bar{i}x}^a\eta^{x0} +\eta^{al}\bar{C}_{\bar{i}l}^0  \right)\langle {\cal{O}}_{i_1} {\cal{O}}_{1,i_2}\dots  {\cal{O}}_{1,i_n}{\cal{O}}_{a}{\cal{O}}_{0}\rangle_{0}  + 
\]
\[
\langle {\cal{O}}_{i_1}{\cal{O}}_{a}\rangle_{0}\frac{1}{2}\left(\bar{C}_{\bar{i}x}^a\eta^{x0}+\eta^{am}\bar{C}_{\bar{i}m}^0 \right)\langle {\cal{O}}_{0}^{\int}{\cal{O}}_{1,i_2}\dots{\cal{O}}_{1,i_n}\rangle_{1} + 
\]
\[
+\sum_{X\cup Y=\{2,n\}} \hspace{-0.3 cm}\langle {\cal{O}}_{i_1}\hspace{-0.3 cm}\prod_{r\in X-\{s\}}\hspace{-0.3 cm}{\cal{O}}_{1,i_r}{\cal{O}}_{i_s}{\cal{O}}_{l}\rangle_{0}\frac{1}{2}\left(\bar{C}_{\bar{i}a}^l\eta^{am} +\eta^{lb}\bar{C}_{\bar{i}b}^m \right)\langle {\cal{O}}_{1,m}\prod_{t\in Y}{\cal{O}}_{1,i_t}\rangle_{1} 
\]

${\cal{O}}_{0}$ can be made disappear from the second line after taking its trivial OPE with any other operator and, being $\eta$ covariantly constant, the second term of the above expression contributes only when the number of operators $n$ is 1. In this case it has the value

\begin{equation}\label{xxx}
\frac{1}{2}\left( \bar{C}_{\bar{i}x}^a\eta^{x0} +\eta^{al}\bar{C}_{\bar{i}l}^0  \right)\eta_{i_1a} = G_{i_1\bar{i}}
\end{equation}

From the other terms instead we obtain

\[
\frac{1}{2}\bar{C}_{\bar{i}}^{lm} \langle {\cal{O}}_{i_1} {\cal{O}}_{1,i_2}\dots  {\cal{O}}_{1,i_n}{\cal{O}}_{l}{\cal{O}}_{m}\rangle_{0} + G_{i_1\bar{i}}\langle {\cal{O}}_{0}^{\int}{\cal{O}}_{1,i_2}\dots{\cal{O}}_{1,i_n}\rangle_{1} 
\]
\[
+
\sum_{X\cup Y=\{2,n\}} \hspace{-0.3 cm}\bar{C}_{\bar{i}}^{lm}\langle {\cal{O}}_{i_1}\hspace{-0.3 cm}\prod_{r\in X-\{s\}}\hspace{-0.3 cm}{\cal{O}}_{1,i_r}{\cal{O}}_{i_s}{\cal{O}}_{l}\rangle_{0}\langle {\cal{O}}_{1,m}\prod_{t\in Y}{\cal{O}}_{1,i_t}\rangle_{1} 
\]

As before 

\[
G_{i_1\bar{i}}\langle {\cal{O}}_{0}^{\int}{\cal{O}}_{1,i_2}\dots{\cal{O}}_{1,i_n}\rangle_{1} = -(n-1)G_{i_1\bar{i}}\langle {\cal{O}}_{1,i_2}\dots{\cal{O}}_{1,i_n}\rangle_{1} 
\]

Again we symmetrize ${\cal{W}}$ by  summing over all the operators to take the role of ${\cal{O}}_{1,i_1}$ and, after the recursion relation has been applied, dividing by the possible choices. The result gives the equivalence of the commutator (\ref{cr1}) applied to $C_{i_1\dots i_n}^1$ and the H.A.E. at genus one with $n\geq2$ marginal operator insertions:

\begin{equation}\label{n2}
\bar{\partial}_{\bar{i}}C_{i_1\dots i_n}^1= [{\cal{W}}_s,\bar{\nabla}_{\bar{i}}]C_{i_1\dots i_n}^1 = 
\end{equation}
\[
=  \frac{1}{2}\bar{C}_{\bar{i}}^{lm}C_{i_1\dots i_nlm}+\bar{C}_{\bar{i}}^{lm}\hspace{-0.3 cm}\sum_{X\cup Y=\{1,n\}}C_{\prod_{r\in X}i_r l}C_{m\prod_{t\in Y}i_t}^1 -(n-1)\sum_{p=1}^{n}G_{i_p\bar{i}}  C_{i_1\dots\hat{p}\dots i_n}^1
\]

The case $n=1$ requires more care. In this situation (\ref{l}) reduces to

\begin{equation}\label{n3}
{\cal{W}}\langle {\cal{O}}_{1,i}\rangle_{1}  = \frac{1}{2}\eta^{\alpha\beta}\langle {\cal{O}}_{i} {\cal{O}}_{\alpha}{\cal{O}}_{\beta}\rangle_{0} + \langle {\cal{O}}_{i}{\cal{O}}_{\alpha}\rangle_{0}\eta^{\alpha\beta}\langle {\cal{O}}_{\beta}^{\int}\rangle_{1}  
\end{equation}

Remembering the additional contribution from (\ref{xxx}) the usual commutator gives the result 

\begin{equation}\label{n4}
\bar{\partial}_{\bar{i}}\langle {\cal{O}}_{1,i}\rangle_{1}  =   [{\cal{W}}_s,\bar{\nabla}_{\bar{i}}]\langle {\cal{O}}_{1,i}\rangle_1=\frac{1}{2}\bar{C}_{\bar{i}}^{lm}C_{ilm}+G_{\bar{i}i}\left( 1+ \langle {\cal{O}}_{0}^{\int}\rangle_{1} \right)  
\end{equation}
 
On a torus without operator insertions the integral of the curvature is zero, thus  the topological twist is trivial and the identification $ {\cal{O}}_{0}=1$ is correct also globally. Then the operator insertion $ {\cal{O}}_{0}^{\int} \equiv \int_{\Sigma}  {\cal{O}}_{0}$ simply multiplies $\langle 1\rangle_{1}$ by the area of the torus, where the latter is the empty genus one amplitude. In \cite{Bershadsky:1993ta} and \cite{Bershadsky:1993cx} \footnote{there is a factor $\frac{1}{2}$ of difference between the two definitions} the operator formulation is used to define $\langle 1\rangle_{1} $ as follows:

\begin{equation}\label{def2}
\langle 1\rangle_{1} = \frac{1}{2}\int \frac{d^2\tau}{\tau_2}Tr[(-1)^F F_L F_R q^{H_L}\bar{q}^{H_R}]
\end{equation}

where $F_{L/R}$ are the left and right fermion number currents. The problem is that this expression is singular for $\tau_2\rightarrow\infty$ and $H_L=H_R=0$, and even worse when multiplied by the area of the torus $\tau_2$; thus it needs regularization and it is uniquely defined only up the addition of a constant ( that drops out only when there is at least one operator ).

So the question of what is the value of $\langle {\cal{O}}_{0}^{\int}\rangle_{1} $ is ill posed, nonetheless there are two natural ways to fix the constant. The first obvious one is to compare (\ref{n4}) to the H.A.E.  for $\langle {\cal{O}}_{1,i}\rangle_{1}$, and obtain

\[
\langle {\cal{O}}_{0}^{\int}\rangle_{1}  =  -\frac{1}{12}Tr(-1)^F
\]

The second method  just generalizes the observation done in the previous section that the operator $ {\cal{O}}_{0}^{\int}$ behaves as minus the dilaton ${\cal{O}}_{1,0}$. Thus we can consider the dilaton equation for the present case, \cite{Verlinde:1990ku}, to get  $\langle {\cal{O}}_{0}^{\int}\rangle_{1}  =  -\frac{1}{12}$; the missing  $Tr(-1)^F$ can be understood from the loop propagation of the topological matter states.

\section{Covariant derivatives}

It is interesting to check explicitly the vanishing of the commutators \footnote{the algebra contains also $ [\nabla_{i},\nabla_j] = [\bar{\nabla}_{\bar{i}},\bar{\nabla}_{\bar{j}}]=0$ but these are not of direct interest to us}

\begin{equation}\label{fc}
[\nabla_{i},\bar{\nabla}_{\bar{j}}] =0
\end{equation}
\[
[{\cal{W}}_s,\nabla_{i}] =0
\]

The second equation is a direct consequence of the U(1) charge condition on the amplitudes: either from (\ref{ccc}) or (\ref{l}) it is immediate to see that neither the action of $D_{i}$ nor of $-2C_i$ permits the appearance of non zero terms. Thus the commutator is satisfied in the trivial way, both ${\cal{W}}_s(\nabla_{i}\dots)$ and $\nabla_{i}({\cal{W}}_s\dots)$ being zero. 

In fact an important difference between the Witten's formulation of the recursion relations (\ref{a}) and (\ref{b}) and the present realization of ${\cal{W}}$, is what does it happen when you act on the recursion relations with an holomorphic derivative that corresponds to the insertion of one more operator, let us say ${\cal{O}}_{i_{n+1}}$.  In \cite{Witten:1989ig}, \cite{Witten:1990hr} it was shown how the action of an holomorphic derivative would transform the recursion relations (\ref{a}) and (\ref{b}) into the corresponding expression for an amplitude containing also ${\cal{O}}_{i_{n+1}}$. This in particular would allow the recursion relations to be valid everywhere in the moduli space of the theory, provided that they are valid in one point and that infinitesimal shifts in the moduli space positions are equivalent to deformations of the action by the same operators associated to the covariant derivative. In our language all this translates to the vanishing of the commutator $[{\cal{W}}_s,D_{i_{n+1}}] =0$. Leaving aside for the moment  the U(1) charge condition that would make both terms of the commutator vanish identically, we can see how the action of $D_{i_{n+1}}$ on the recursion relation generated by ${\cal{W}}$ is different by the recursion relation for an amplitude with one additional operator insertion . For example at genus one we have 
 
\begin{equation}\label{q1}
D_{i_{n+1}}{\cal{W}} \langle {\cal{O}}_{1,i_1}\dots{\cal{O}}_{1,i_n}\rangle_{1}= 
\end{equation}
\[
=D_{i_{n+1}}(\frac{1}{2}\eta^{\alpha\beta}\langle {\cal{O}}_{i_1}{\cal{O}}_{1,i_2}\dots{\cal{O}}_{1,i_n}{\cal{O}}_{\alpha}{\cal{O}}_{\beta}\rangle_{0}) + 
D_{i_{n+1}}(\langle {\cal{O}}_{i_1}{\cal{O}}_{\alpha}\rangle_{0}\eta^{\alpha\beta}\langle {\cal{O}}_{\beta}^{\int}{\cal{O}}_{1,i_2}\dots{\cal{O}}_{1,i_n}\rangle_{1}  ) +
\]
\[
+ D_{i_{n+1}}(\langle {\cal{O}}_{i_1}\prod_{r\in X-\{s\}}{\cal{O}}_{1,i_r}{\cal{O}}_{i_s}{\cal{O}}_{\alpha}\rangle_{0}\eta^{\alpha\beta}\langle {\cal{O}}_{1,\beta}\prod_{t\in Y}{\cal{O}}_{1,i_t}\rangle_{1}  ) = 
 \]

\[= \frac{1}{2}\eta^{\alpha\beta}\langle {\cal{O}}_{i_1}{\cal{O}}_{1,i_2}\dots{\cal{O}}_{1,i_{n+1}}{\cal{O}}_{\alpha}{\cal{O}}_{\beta}\rangle_{0} + \langle {\cal{O}}_{i_1}{\cal{O}}_{\alpha}\rangle_{0}\eta^{\alpha\beta}\langle {\cal{O}}_{\beta}^{\int}{\cal{O}}_{1,i_2}\dots{\cal{O}}_{1,i_{n+1}}\rangle_{1}  + 
\]
\[
+\langle {\cal{O}}_{i_1}\prod_{r\in X-\{s\}}{\cal{O}}_{1,i_r}{\cal{O}}_{i_s}{\cal{O}}_{\alpha}{\cal{O}}_{1,i_{n+1}}\rangle_{0}\eta^{\alpha\beta}\langle {\cal{O}}_{1,\beta}\prod_{t\in Y}{\cal{O}}_{1,i_t}\rangle_{1}   + 
\]
\[
+\langle {\cal{O}}_{i_1}\prod_{r\in X-\{s\}}{\cal{O}}_{1,i_r}{\cal{O}}_{i_s}{\cal{O}}_{\alpha}\rangle_{0}\eta^{\alpha\beta}\langle {\cal{O}}_{1,\beta}\prod_{t\in Y}{\cal{O}}_{1,i_t}{\cal{O}}_{1,i_{n+1}}\rangle_{1}   
\]

This fails to reproduce the recursion relation for a genus one amplitude with $n+1$ operator insertions, ${\cal{W}} \left(D_{i_{n+1}}\langle {\cal{O}}_{1,i_1}\dots{\cal{O}}_{1,i_n}\rangle_{1}\right)$, as it misses the term

\begin{equation}\label{q66}
\langle {\cal{O}}_{i_1}{\cal{O}}_{i_{n+1}}{\cal{O}}_{\alpha}\rangle_{0}\eta^{\alpha\beta}\langle {\cal{O}}_{1,i_2}\dots{\cal{O}}_{1,i_n}{\cal{O}}_{1,\beta}\rangle_{1}   
\end{equation}

 which would come as the special case $X=\{s\}=\{n+1\}$ from the last term in equation (\ref{l}), and  cannot be obtained by the action of $D_{i_{n+1}}$ on any term inside ${\cal{W}} \langle {\cal{O}}_{1,i_1}\dots{\cal{O}}_{1,i_n}\rangle_{1}$.
 
 However this is somehow irrelevant as the right hand side of both (\ref{ccc}) and (\ref{l}) is always zero due to the U(1) charge condition, and $D_{i_{n+1}}$ does not change the charge. A better question would come from considering not $[{\cal{W}}_s,D_{i_{n+1}}] $ but instead $[[{\cal{W}}_s,\bar{\nabla}_{\bar{i}}] ,D_{i_{n+1}}] $, that is switching from the action of ${\cal{W}}_s$ alone to the commutator $ [{\cal{W}}_s,\bar{\nabla}_{\bar{i}}]$, which applied to the amplitudes produces a nonzero result, and evaluating its commutator with $D_{i_{n+1}}$. Having $[{\cal{W}}_s,\bar{\nabla}_{\bar{i}}] = \bar{\nabla}_{\bar{i}}$ we are naturally brought to the direct analysis of $[\bar{\nabla}_{\bar{i}},D_{i_{n+1}}]$, or equivalently of $[\nabla_{i_{n+1}},\bar{\nabla}_{\bar{i}}]$ ( known the latter we can compute the former ). This can be generically applied to amplitudes $C^{g}_{i_1\dots i_n}$. Its zero result is a well known direct consequence of the tt* equations, but it can be rederived as well from the explicit form of the H.A.E. after it is rewritten as \footnote{this computation in fact requires some care as $C^{g}_{i_1\dots i_n}$ is a section of ${\cal{L}}^{2-2g-n}\otimes Sym TM^{n}$, the indexes $i_1\dots i_n$ belonging to the latter while ${\cal{L}}^{2-2g-n}$ can be formally represented by the identity operator $0$ on which both $C_{i_{n+1}}$ and $\bar{C}_{\bar{i}}$ should be made acting ( even if the second is in fact vanishing again due to the U(1) anomaly condition ).}

\[
[D_{i_{n+1}},\bar{\partial}_{\bar{i}}]C^{g}_{i_1\dots i_n} = -[C_{i_{n+1}},\bar{C}_{\bar{i}}]C^{g}_{i_1\dots i_n}
\]

The fact that the H.A.E. for amplitudes with operator insertions is the direct consequence of the H.A.E. for amplitudes without operators and the first commutator in (\ref{fc}), was already well known since \cite{Bershadsky:1993cx}. Here we are proposing an enlarged algebra based on the definition of the operator ${\cal{W}}_s$ whose consequence, without further input, is the H.A.E. for generic amplitudes with operators \footnote{ so far only up to genus one }. Moreover, and this is the central point, this algebra makes direct contact with the Witten derivation of the topological recursion relations.

\section{Higher genus}
 
We want to generalize this construction at higher genus. Because the derivation used by Witten does not straightforwardly extend for generic genus $g$ ( though recursion relations for Gromov - Witten invariants have been written at higher genus, \cite{Eguchi:1998ji}
  and \cite{Gathmann} ),  we reverse the procedure used so far. First we define ${\cal{W}}$ in order to reproduce the  form of the H.A.E., and then we attempt an algebraic interpretation to make contact with the Witten's derivation of the topological recursion relations. Following this philosophy we divide ${\cal{W}}$ in two parts and define ${\cal{W}}C^g_{i_1 \dots i_n} = \left({\cal{W}}_1 + {\cal{W}}_2\right)  C^g_{i_1 \dots i_n}$ as 
  
 \begin{equation}\label{r}
{\cal{W}}_1\langle {\cal{O}}_{1,i_{1}}\dots{\cal{O}}_{1,i_{n}}\rangle_{g} = \sum_{X\cup Y=\{2\dots n \}} \langle {\cal{O}}_{i_{1}}\prod_{r\in X-\{s\}}{\cal{O}}_{1,i_{r}}{\cal{O}}_{i_{s}}{\cal{O}}_{\alpha}\rangle_{0}\eta^{\alpha\beta}\langle {\cal{O}}_{1,\beta}\prod_{t\in Y}{\cal{O}}_{1,i_{t}}\rangle_{g} + 
\end{equation}
\[
+ \langle {\cal{O}}_{i_{1}}{\cal{O}}_{\alpha}\rangle_{0}\eta^{\alpha\beta}\langle {\cal{O}}_{\beta}^{\int}{\cal{O}}_{1,i_{2}}\dots {\cal{O}}_{1,i_{n}}\rangle_{g}  
\]

\begin{equation}\label{r2}
{\cal{W}}_2\langle {\cal{O}}_{1,i_{1}}\dots{\cal{O}}_{1,i_{n}}\rangle_{g} = \frac{1}{2}\eta^{\alpha\beta}\langle {\cal{O}}_{1,i_{1}}\dots{\cal{O}}_{1,i_{n}}{\cal{O}}_{1,\alpha}{\cal{O}}_{1,\beta}\rangle_{g-1} +
\end{equation}
\[
+ \frac{1}{2}\sum_{X\cup Y=\{1\dots n \}} \sum_{f= 1}^{g-1}\langle \prod_{r\in X}{\cal{O}}_{1,i_{r}}{\cal{O}}_{1,\alpha}\rangle_{g-r}\eta^{\alpha\beta}\langle {\cal{O}}_{1,\beta}\prod_{t\in Y}{\cal{O}}_{1,i_{t}}\rangle_{f}  
\]

\begin{figure}[htb]
 \centering
  \vspace{-0pt}
 \def\svgwidth{280pt}
  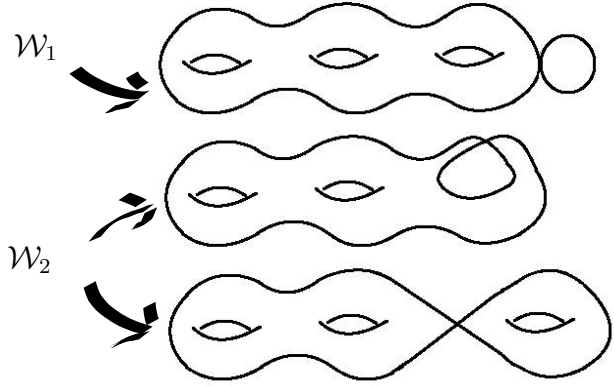
  \vspace{-0pt}  
  \caption{ The action of ${\cal{W}}_1$ and  ${\cal{W}}_2$ is represented. }
  \label{figura2} 
\end{figure}

The two operators ${\cal{W}}_1$ and  ${\cal{W}}_2$ correspond pictorially to the contributions represented in figure (\ref{figura2}); note the novelty of the terms dividing the surface into a genus $f$ and $g-f$ pair, when a dividing cycle shrinks forming a dividing node. 

From equation (\ref{r}) we are again led to interpret ${\cal{W}}_1$ as the evaluation of the zeros of a section associated to the cohomology class of ${\cal{O}}_{1,i_{1}}$, as it transforms to ${\cal{O}}_{1,i_{1}}\rightarrow {\cal{O}}_{i_{1}}$ . So the first question is why also ${\cal{W}}_2$ appears? And the second is what is the physical object inside the correlation function associated to the two form we presume has been integrated over to pass to the right hand side of (\ref{r2})?

Both answers come from the generalization of the procedure of symmetrization we have used to pass from ${\cal{W}}$ to ${\cal{W}}_s$. The reason we should include also ${\cal{W}}_2$ in the game is that for genus $g\geq 2$ amplitudes the spin two supercurrents $G^-,\bar{G}^-$ appear not only inside the operator insertions but also in  the topological measure $\prod_{b=1}^{3g-3}(G^{-},\mu_{b})(\bar{G}^{-},\bar{\mu}_{\bar{b} })$, which is an essential ingredient in the definition (\ref{c1tt}). Thus the process of symmetrization enlarges from summing other all the operators to summing other everything that contains $G^-,\bar{G}^-$; when $G^-,\bar{G}^-$ belong to the operators we are led to ${\cal{W}}_1$, when they belong to the measure we are led to ${\cal{W}}_2$. More precisely we identify the topological measure as the physical object associated to the two dimensional differential form we are supposed to integrate in order to reproduce the right hand side of (\ref{r2}). In order to predict the form of ${\cal{W}}_2$ we would need to  find an appropriate section with zeros supported only at the boundary of ${\cal{M}}_{g,n}$, and then integrate. But it is probably too complicated. However an indirect result can be inferred, as we have done, starting from the H.A.E. and giving the form of ${\cal{W}}_2$ which is compatible with it.

 In fact we can check that, passing from left to right in equation (\ref{r2}), the total topological measure always looses three of the terms $(G^{-},\mu_{b})(\bar{G}^{-},\bar{\mu}_{\bar{b} })$, two of which come back as the measure associated to the moduli parametrizing the position of the node and thus encircling the operators ${\cal{O}}_{\alpha}$ and ${\cal{O}}_{\beta}$, as well known from the construction in \cite{Bershadsky:1993cx}.

Accepted the result the evaluation of the commutator $ [{\cal{W}}_s,\bar{\nabla}_{\bar{i}}] = \bar{\nabla}_{\bar{i}}$ proceeds analogously as before, symmetrizing ${\cal{W}}_1$ ( ${\cal{W}}_2$ is already defined as symmetrized over all the $3g-3$ contributions from $(G^{-},\mu_{b})(\bar{G}^{-},\bar{\mu}_{\bar{b} })$ ), and it produces the most general form of the H.A.E., for correlation functions at genus $g\geq 2$ and $n$ marginal operators.

\section{Conclusions}

In this work we have established a connection between the topological recursion relations and the H.A.E. . In detail we have first defined an operator ${\cal{W}}_s$ mimicking the symmetrization of what would be the action of the topological recursion relations on BCOV type amplitudes, formally replacing the role of the scalar $\phi$ with $G^-,\bar{G}^-$. Then we have written an algebra containing ${\cal{W}}_s$, $\nabla_i$ and $\bar{\nabla}_{\bar{i}}$ that introduces the antiholomorphic information contained into the topological string amplitudes ( expected to satisfy the recursion relations only in the holomorphic limit ) and generalizes the tt* equations. Its validity is confirmed by being equivalent to the H.A.E. when applied to topological string amplitudes with marginal operator insertions. Thus the connection between the formalism developed by Witten and the H.A.E. of Bershadsky, Cecotti, Ooguri and Vafa.

A first direction for future work is certainly a deeper understanding of the above construction and a natural guess is the possibility of a purely geometrically algebraic derivation of the H.A.E. . Further it follows the question if it exists some deeper understanding for the construction of the operator ${\cal{W}}_s$ ( counterpart of the possible geometrically algebraic derivation in the holomorphic limit ) and of the algebra (\ref{cr1}), (\ref{cr2}) and (\ref{cr3}) ( that would correspond to the antiholomorphic information ). Indeed all of this can be of conceptual interest as the H.A.E. is already well known to have deep and interesting physical interpretations beyond the usual worldsheet construction ( see for example \cite{Witten:1993ed} ).

As a more specific subproblem it would also be interesting to understand the process of symmetrization of the operator ${\cal{W}}_s$ that has not a clear meaning from an algebraic point of view. For amplitudes at genus zero and one it looks only as an overall normalization in front of the recursion relations, but for higher genus it is the base for the construction of  ${\cal{W}}_1$ and ${\cal{W}}_2$, as we have seen.

 Besides the result per se we would also like to look for some practical applications.  A first idea would be to consider our algebraic construction as an initial step into proving the integrability for generic topological string amplitudes, extending the result already achieved for solutions of the tt* equations. This is certainly worth of future work. Moreover it is plausible that the present construction can be an important tool for developing new techniques for solving the H.A.E. in a perhaps more efficient way.

\subsection*{Acknowledgments}

It is a pleasure to thank Giulio Bonelli, Tohru Eguchi, Kazuhiro Sakai, Tadashi Takayanagi and Alessandro Tanzini  for various discussions and useful comments. This work was supported by the Japanese Society for the Promotion of Science (JSPS).


\begin{thebibliography}{20}

\bibitem{Bershadsky:1993ta}
  M.~Bershadsky, S.~Cecotti, H.~Ooguri and C.~Vafa,
  ``Holomorphic anomalies in topological field theories,''
  Nucl.\ Phys.\  B {\bf 405} (1993) 279
  [arXiv:hep-th/9302103].
  
\bibitem{Bershadsky:1993cx}
  M.~Bershadsky, S.~Cecotti, H.~Ooguri and C.~Vafa,
  ``Kodaira-Spencer theory of gravity and exact results for quantum string amplitudes,''
  Commun.\ Math.\ Phys.\  {\bf 165} (1994) 311
  [hep-th/9309140].
  
\bibitem{Cecotti:1991me}
  S.~Cecotti and C.~Vafa,
  ``Topological antitopological fusion,''
  Nucl.\ Phys.\ B {\bf 367} (1991) 359.
  
\bibitem{Eguchi:1998ji}
  T.~Eguchi and C.~-S.~Xiong,
  ``Quantum cohomology at higher genus: Topological recursion relations and Virasoro conditions,''
  Adv.\ Theor.\ Math.\ Phys.\  {\bf 2} (1998) 219
  [hep-th/9801010].
 
\bibitem{Gathmann}
  A.~Gathmann,
  ``Topological recursion relations and Gromov-Witten invariants in higher genus,''
  [arxiv:math/0305361].
  
\bibitem{Gomez:1994qk}
  C.~Gomez and E.~Lopez,
  ``On the algebraic structure of the holomorphic anomaly for N=2 topological strings,''
  Phys.\ Lett.\ B {\bf 334} (1994) 323
  [hep-th/9405071].

\bibitem{Verlinde:1990ku}
  E.~P.~Verlinde and H.~L.~Verlinde,
  Nucl.\ Phys.\ B {\bf 348} (1991) 457.
  
\bibitem{Witten:1989ig}
  E.~Witten,
  ``On The Structure Of The Topological Phase Of Two-dimensional Gravity,''
  Nucl.\ Phys.\ B {\bf 340} (1990) 281.
      
\bibitem{Witten:1990hr}
  E.~Witten,
  ``Two-dimensional gravity and intersection theory on moduli space,''
  Surveys Diff.\ Geom.\  {\bf 1} (1991) 243.      
     
\bibitem{Witten:1991zz}
  E.~Witten,
  ``Mirror manifolds and topological field theory,''
  In *Yau, S.T. (ed.): Mirror symmetry I* 121-160
  [hep-th/9112056].

\bibitem{Witten:1993ed}
  E.~Witten,
  ``Quantum background independence in string theory,''
  hep-th/9306122.
    
\end{thebibliography}
\end{document}